\documentclass[twocolumn,showpacs]{revtex4}
\usepackage{amsmath}
\usepackage{amssymb}
\usepackage{bm}
\usepackage{epsfig}
\usepackage{graphicx}

\begin{document}

\title{Spin-orbit effect on electron-electron interaction and fine structure of electron complexes in quantum dots}

\author{M.M. Glazov}

\affiliation{Ioffe Physical-Technical Institute RAS, 194021 St.-Petersburg, Russia}

\author{V.D. Kulakovskii}

\affiliation{Institute of Solid State Physics RAS, 142432 Chernogolovka, Russia}


\begin{abstract}
Spin-orbit effects on electron-electron interaction are studied theoretically. The corrections to the Coulomb interaction of quantum well electrons induced by the spin-orbit coupling are derived. The developed theory is applied to calculate the energy spectrum fine structure of an electron pair triplet states localized in small lateral disk-shaped quantum dots. We show that the spin degeneracy of a triplet state is completely lifted in anisotropic quantum dots. Isotropic quantum dots also demonstrate peculiar fine structure of triplet states caused by the spin-orbit interaction.
\end{abstract}

\date{\today}

\maketitle

\section{Introduction}

Spin-dependent phenomena in semiconductors are actively studied nowadays.~\cite{dyakonov_book} Quantum dots demonstrate a rich variety of promising properties, among those are long spin relaxation times of single electrons or holes and fast spin dynamics of electron-hole complexes.~\cite{flissikowski01,Ignatiev05} Spin effects are usually related with the fine structure of energy spectrum of charge carriers and their complexes. Thus, understanding the spin-dependent fine structure of electronic states is of high importance~\cite{ivchenko05a}.

Due to the time-reversal symmetry the single electron states in quantum dots are spin-degenerate and form Kramers doublets. To the contrary, the localized states of an electron-hole pair (zero dimensional excitons) are characterized by an integer value of the total spin and demonstrate diverse fine structure caused by the short- and long-range exchange interactions between an electron and a hole.~\cite{goupalov98,maialle00} Such fine structure is extensively studied theoretically and experimentally, see Ref.~\cite{ivchenko05a} and references therein. Recently, an increasing interest is attracted by two-electron complexes in single quantum dots or in quantum dot molecules which are probed by the photoluminescence spectroscopy.~\cite{akimov05,kul1:07,kul2:07eng} 
Two electron states may serve as final states in the processes of the radiative recombination in photoexcited doubly charged quantum dots.

The ground state of a pair of electrons is a spin singlet due to the Pauli principle. In the absence of the spin-orbit interaction, the excited triplet states are degenerate as well. An interference of the spin-orbit and Coulomb interactions can, in principle, lift this degeneracy.

We demonstrate here that this is exactly the case: in anisotropic quantum dots the triplet states of electron pairs are split into three fine sublevels, one of which corresponds to zero projection of the total spin on the growth axis and two others correspond to linear combinations of the states with total spin projections being $\pm 1$ similar to those of localized excitons.

The microscopic origin of the two electron energy spectrum fine structure is the spin-orbit interaction. At first, it results in the wavevector dependent splitting of the free electron energy spectrum.
The effect of conduction band spin splitting caused by the absence of an inversion center in the system on the exchange interaction of localized electrons has been analyzed in Refs.~\cite{kkavokin01,kkavokin04}, see also Ref.~\cite{gangadharaiah:156402}. 
It was demonstrated that the spin degeneracy of the triplet two electron states is not removed up to a high (fourth) order in the conduction band splitting.~\cite{gangadharaiah:156402}

However, the spin-orbit interaction can manifest itself not only in the splitting of electronic dispersion, but also in the modification of electron scattering.~\cite{abakumov72}
A possibility of the spin-flip at electron-electron or hole-hole collision has been discussed in Refs.~\cite{boguslawski,schneider:420,badescu:161304}. The effect of the spin-orbit corrections to the Coulomb interaction on the asymmetric exchange interaction of electrons was discussed in Ref.~\cite{badescu:161304}. The fine structure of a triplet state of localized electron pair in a quantum dot due to the spin-orbit induced corrections to the Coulomb interaction has not been addressed yet to the best of our knowledge. The present paper aims to fill this apparent gap.

In the present paper we derive within the framework of Kane model the spin-orbit coupling induced corrections to the Coulomb interaction between two electrons. The developed formalism is applied to calculate the fine structure of the triplet state of a two electron complex localized in a quantum disk. The splitting of a triplet state is obtained in the second order in the spin-orbit interaction and in the first order in the Coulomb repulsion. We analyze the effects of disk geometry on the fine structure of a pair of electrons.

\section{Spin-orbit terms in electron-electron interaction}

Spin-orbit interaction in semiconductors is largely determined by the spin-orbit splitting of the valence band. Its effect on the conduction band states can be described by $\bm k \cdot \bm p$ admixture of the valence band states. Our ultimate goal is to obtain the fine structure of a two-electron complex in a quantum disk (lateral quantum dot), i.e. in the system where the quantization along the growth axis is much stronger as compared with the lateral quantization. To do so, we firstly derive spin-orbit interaction induced corrections to the Coulomb potential of quantum well electrons interaction and secondly we apply the obtained effective interaction to determine the energy spectrum fine structure of two electrons localized in the disk.

The direct band III-V semiconductor structure is considered. We use the 8-band Kane model in order to describe the spin-orbit contributions to the electron-electron interaction. This model disregards an absence of an inversion center in the bulk material. The effects of bulk inversion asymmetry on electron-electron interaction require a separate study. As we see below, in anisotropic quantum disks the triplet state degeneracy is completely lifted therefore bulk inversion asymmetry may lead only to the quantitative modifications of the spectrum.
 
As a starting point it is convenient to represent a free electron wavefunction in a quantum well grown along $z\parallel [001]$ axis in the fist order of $\bm k \cdot \bm p$ interaction as~\cite{ivchenko05a,suris86,averkiev02}
\begin{equation}
 \label{kane}
\Psi_{s,\bm k}(\bm \rho,z) = e^{\mathrm i \bm k\bm \rho} [ \mathsf S_{\bm r} + \mathrm i {\mathbf R}_{\bm r} \cdot (A \hat{\bm K} - \mathrm i B \hat{\bm \sigma} \times \hat{\bm K})] \varphi(z) \, |  \chi_s\rangle.
\end{equation}
Here $\bm r = (\bm \rho,z)$ is the electron position vector,  $\hat{\bm K} = (\bm k, -\mathrm i \partial/\partial z)$ is the wave vector of the electron, $\varphi(z)$ is the smooth envelope describing its size quantization along $z$ axis, $\hat{\bm \sigma}$ is the electron spin operator, $\mathsf S_{\bm r}$ and $\mathbf R_{\bm r} = (\mathsf X_{\bm r}, \mathsf Y_{\bm r}, \mathsf Z_{\bm r})$ are $s$-type and $p$-type Bloch functions taken at the $\Gamma$ point, $|\chi_s\rangle$ is a spinor. Constants $A$ and $B$ are equal to
\begin{gather}\label{AB}
 A = \frac{\mathrm i \hbar p_{cv}}{3m_0}\left(\frac{2}{E_g} + \frac{1}{E_g + \Delta}\right), \\ B = -\frac{\mathrm i \hbar p_{cv}}{3m_0}\left(\frac{1}{E_g} - \frac{1}{E_g + \Delta}\right),\nonumber
\end{gather}
where $E_g$ and $\Delta$ are the band gaps $\Gamma_6$---$\Gamma_8$ and $\Gamma_8$---$\Gamma_7$, respectively, $p_{cv}$ is the interband momentum matrix element, and $m_0$ is the free electron mass. Normalization constant in Eq.~\eqref{kane} is omitted. Equation~\eqref{kane} is valid provided that the electron energy referred to the conduction band bottom is much smaller as compared to $E_g$ and $\Delta$.

The Hamiltonian describing Coulomb scattering of two electrons from the states $(\bm k s, \bm k's')$ to the states $(\bm p s_1, \bm p's_1')$ taking into account their indistinguishability and allowing for the spin-orbit interaction is formed from the matrix elements of bare Coulomb potential 
\[
V(\bm r_1 - \bm r_2) = \frac{e^2}{\varkappa |\bm r_1 - \bm r_2|}, 
\]
taken between the Slater determinants composed of multicomponent functions Eq.~\eqref{kane}. Here $e$ is the elementary charge and $\varkappa$ is the static dielectric constant. We consider the spin-orbit interaction as a small perturbation, therefore it is enough to calculate the matrix element of $V(\bm r_1 - \bm r_2)$ for non-symmetrized wavefunction and antisymmetrize it afterwards. The non-symmetrized matrix element reads
\begin{widetext}
\begin{multline}
 \label{Coulomb1}
M(\bm ks, \bm k's' \to \bm ps_1, \bm p's_1') = \frac{2\pi e^2}{\Xi \varkappa q} \delta_{\bm k+\bm k',\bm p + \bm p'}\langle \chi_{s_1}\chi_{s_1'} |  \left\{ F_{22}^{00}(q) + \right.\\
 \xi \left[ [\hat{\bm \sigma}^{(1)} \times (\bm p+\bm k)]_z F_{12}^{10}(q)  +  [\hat{\bm \sigma}^{(2)} \times (\bm p'+\bm k')]_z F_{21}^{01}(q) \right. +
\\ \left.\mathrm i F_{22}^{00}(q) \left([\bm p \times \bm k]\hat{\bm \sigma}^{(1)} + [\bm p' \times \bm k']\hat{\bm \sigma}^{(2)}\right)\right]+
 \xi^2\left[[\hat{\bm \sigma}^{(1)} \times (\bm p+\bm k)]_z [\hat{\bm \sigma}^{(2)} \times (\bm p'+\bm k')]_z F_{11}^{11}(q) + \right.
\\ \mathrm i [\hat{\bm \sigma}^{(1)} \times (\bm p+\bm k)]_z [\bm p'\times\bm k']_z\hat{\sigma}^{(2)}_z F_{12}^{10}(q) + \mathrm i [\bm p \times \bm k]_z\hat{\sigma}^{(1)}_z [\hat{\bm \sigma}^{(2)}\times (\bm p' + \bm k')]_z F_{21}^{01}(q)-\\
 \left.\left.([\bm p \times \bm k]\hat{\bm \sigma}^{(1)} )([\bm p' \times \bm k']\hat{\bm \sigma}^{(2)}) F_{22}^{00}(q) \right]\right\}| \chi_{s}\chi_{s'}\rangle.
\end{multline}
\end{widetext}
Here $\Xi$ is the normalization area, $\bm q = \bm p - \bm k$ is the transferred wavevector, the parameter $\xi = 2AB+B^2$ characterizes the strength of the spin-orbit interaction, $\hat{\bm \sigma}^{(1)}$ and $\hat{\bm \sigma}^{(2)}$ are the spin operators of the first and second electrons (acting on spinors $|\chi_s\rangle$,$|\chi_{s_1}\rangle$ and on  $|\chi_{s'}\rangle$,$|\chi_{s_1'}\rangle$, respectively). Form-factors $F_{ij}^{kl}(q)$ depend on the shape of the electron envelope in the growth direction, 
\begin{equation}
F_{ij}^{kl}(q) = 
\int \mathrm dz_1\mathrm dz_2 e^{-q|z_1 - z_2|} [\varphi(z_{1})]^i [\varphi(z_2)]^j [\varphi'(z_1)]^k[\varphi'(z_2)]^l.\nonumber
\end{equation}
 
In Eq.~\eqref{Coulomb1} we have neglected the valence band admixture induced corrections to the spin-independent part of Coulomb matrix element along with quadratic in wavevectors corrections to the linear in $\xi$ terms. 
Terms linear in wavevectors in Eq.~\eqref{Coulomb1} (cf.~\cite{badescu:161304}) describe skew electron-electron scattering and have the same form as structure-inversion asymmetry induced electron-impurity or electron phonon scattering terms, see Refs.~\cite{averkiev02,tarasenko05} and references therein. They may arise only in quantum wells with structure inversion asymmetry. The terms linear in $\xi$ and quadratic in the wavevectors contain cross-products of $\bm k$, $\bm p$ and of $\bm k'$, $\bm p'$ and describe Mott effect at electron-electron scattering.~\cite{boguslawski} Terms quadratic in $\xi$ describe spin-spin interaction and are responsible for the mutual spin-flip of electrons. These terms are of importance in determination of fine structure of a pair of localized carriers.

\section{Fine structure of a two-electron complex}

\begin{figure}
 \includegraphics[width=0.27\linewidth]{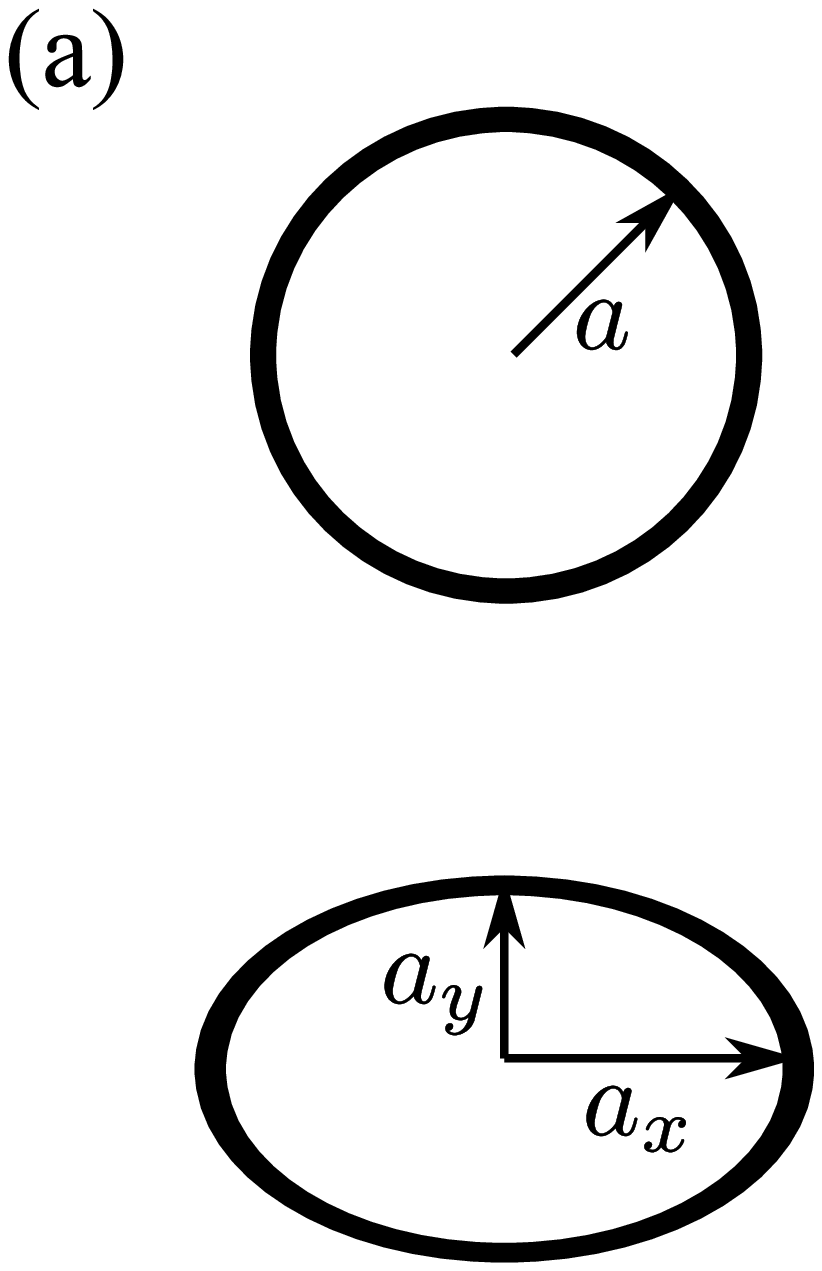}\includegraphics[width=0.69\linewidth]{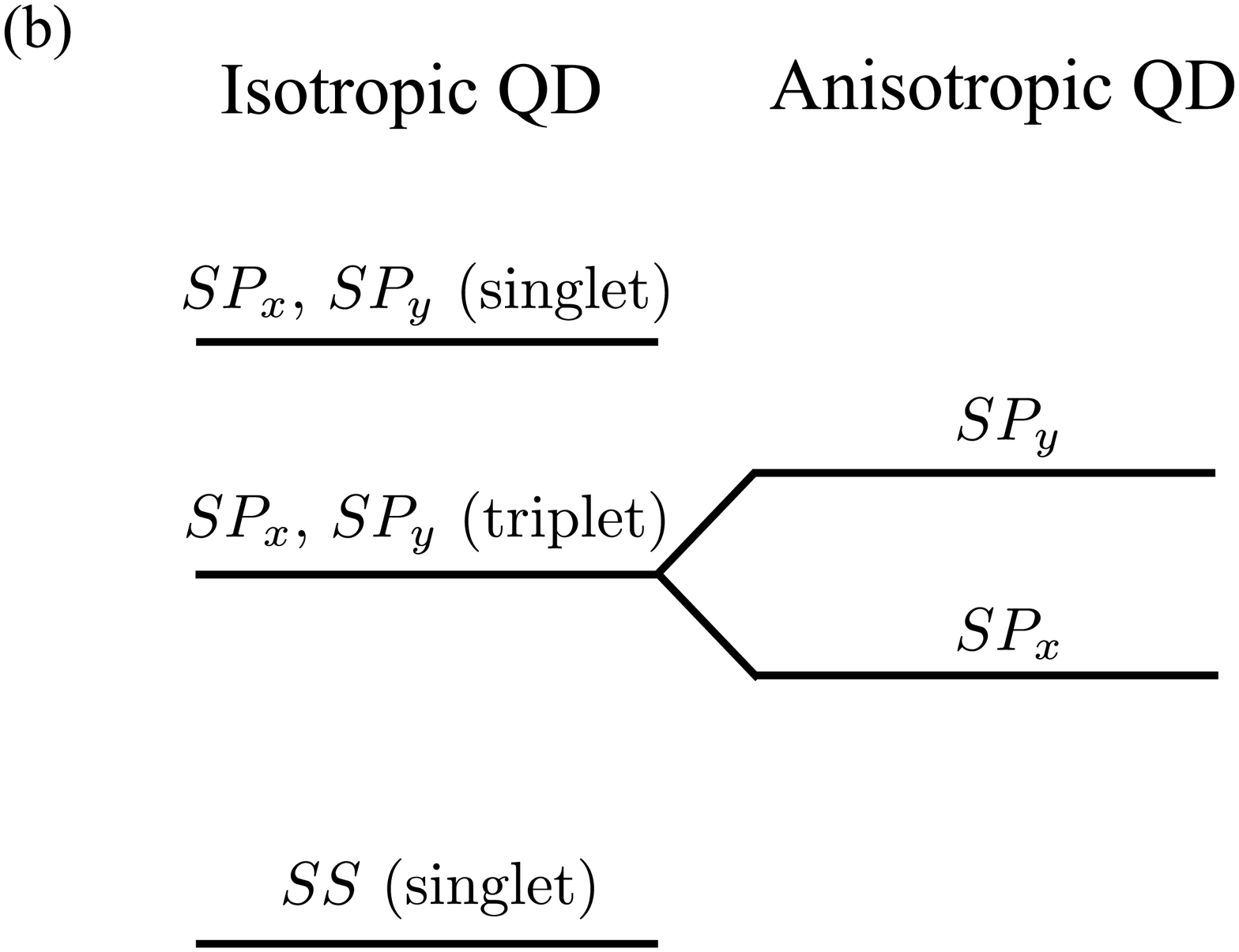}
 \caption{
Panel (a): axially symmetric and anisotropic quantum disks (top view). Panel (b): schematic illustration of two electron orbital states in quantum disks. Energy gaps are shown not to scale. The splitting of $SP$ triplet state due to a possible lateral potential anisotropy is shown. The singlet $SP$ orbital state is also split by a disk anisotropy in the same fashion (not shown).
}\label{fig:orb}
\end{figure}

We consider a quantum dot embedded into a quantum well. It is assumed that the quantum dot radius $a$ or effective radii $a_x$, $a_y$ along the main in plane axes of the quantum dot, $x$, $y$, exceed by far the quantum well width (i.e. the dot height) $d$, see Fig.~\ref{fig:orb}(a). On the other hand, the lateral size of the quantum dot is supposed to be small enough in order to allow us to fix electron envelope functions and treat Coulomb interaction between charge carriers, Eq.~\eqref{Coulomb1}, as a small perturbation. Thus, the following energy hierarchy is assumed: the largest energy scale is the size-quantization along $z$ axis, then lateral size-quantization, thereupon the Coulomb interaction without spin-orbit terms and, finally, the spin-orbit induced corrections to the electron-electron interaction.

The structure of orbital states is schematically shown in Fig.~\ref{fig:orb}(b). The ground state of two electrons is a spin singlet (with total spin equal to $0$), their orbital functions are the same and belong to the lowest size quantization level of the disk [$S$-states, described by single electron lateral envelope $S(\bm \rho)$]. This state is denoted as $SS$-orbital state. The spin-orbit interaction shifts slightly the ground state position, this effect is ignored hereafter.

We focus on the lowest excited states, namely, those formed of one electron in the ground state and another one in the first excited $P$-orbital state. There are two such states, $SP_x$ and $SP_y$ (with node on $y$- and $x$-axis, respectively), described by the lateral envelope functions $P_x(\bm \rho)$ and $P_y(\bm \rho)$. In axially symmetric disks these states are degenerate, they can be split due to an anisotropy of the lateral potential. We first consider the case of anisotropic disk where these states are independent, and further, the case of axially symmetric disk is discussed. In the absence of the spin-orbit interaction each $SP_i$ state ($i=x,y$) is split in spin singlet and triplet by the electron-electron exchange interaction. Singlet states are disregarded in what follows.

\subsection{Symmetry arguments}

Before presenting a microscopic theory let us analyze symmetry restrictions on the fine structure of the triplet states. Following the method of invariants we construct the spin-dependent part of the electron-electron interaction Hamiltonian for a given triplet state $SP_i$ ($i=x$ or $y$) from the operators of the total momentum $1$, $\hat S_\alpha$ ($\alpha=x,y,z$). Time reversal symmetry implies that only quadratic combinations of $\hat S_\alpha$ enter the spin Hamiltonian. In anisotropic disks with $x$ and $y$ being the main axes $\hat S_x^2$ and $\hat S_y^2$ are invariant, therefore, the effective Hamiltonian
can be recast as
\begin{equation}
 \label{H:sym:ii}
\hat{\Delta}_{ii} =  \mathcal A_i \hat S_x^2 + \mathcal B_i \hat S_y^2 - (\mathcal A_i + \mathcal B_i) \hat S_z^2,
\end{equation}
where $\mathcal A_i$ and $\mathcal B_i$ are some constants. The term proportional to $\hat S_z^2$ is added in order to eliminate the total energy shift of the triplet. Hence, a fine structure of each triplet orbital is determined by two linearly-independent parameters. 

\begin{figure}
\includegraphics[width=0.8\linewidth]{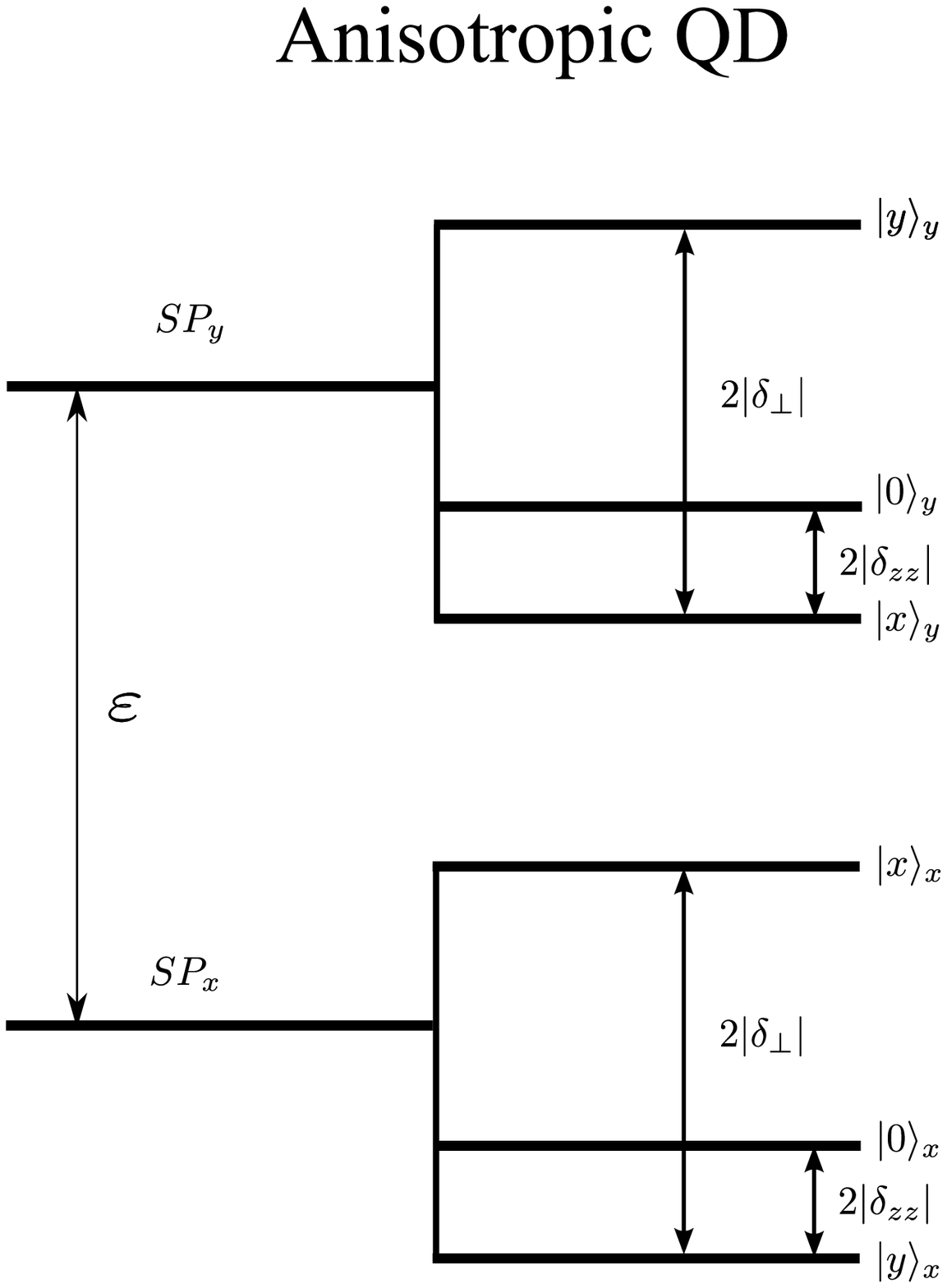}
\caption{Schematic illustration of the fine structure of triplet states in an anisotropic quantum disk. $SP_x$ and $SP_y$ denote orbital states $\Psi_x(\bm \rho_1,\bm \rho_2)$ and $\Psi_y(\bm \rho_1,\bm \rho_2)$ [Eq.~\eqref{psifun}], respectively, being splitted due to the lateral anisotropy. Here $|0\rangle$, $|x\rangle$ and $|y\rangle$ denote spin states (i.e. those with the total spin $z$ projection $m_z$ being $0$, and two linear combinations of $m_z=\pm 1$ states). Subscripts $x$ and $y$ denote orbital states.}
\label{fig:aniso} 
\end{figure}

In a general case of arbitrary $\mathcal A_i$ and $\mathcal B_i$ each orbital state is split into three sublevels, according to the $z$-projection of the total spin, $m_z$: the state with $m_z=0$ and two states being the combinations of $m_z=\pm 1$ states, see Fig.~\ref{fig:aniso}. The latter form ``linearly-polarized'' combinations:
$|x\rangle = (|+1\rangle+|-1\rangle)/\sqrt{2}$, $|y\rangle = -\mathrm i(|+1\rangle-|-1\rangle)/\sqrt{2}$, similar to those of heavy-hole exciton in an anisotropic quantum disk.~\cite{ivchenko05a,goupalov98,glazov2007a} Such a fine structure is specific to the \emph{anisotropic} quantum disks where $SP_x$ and $SP_y$ orbitals are split in energy.

\begin{figure}
\includegraphics[width=\linewidth]{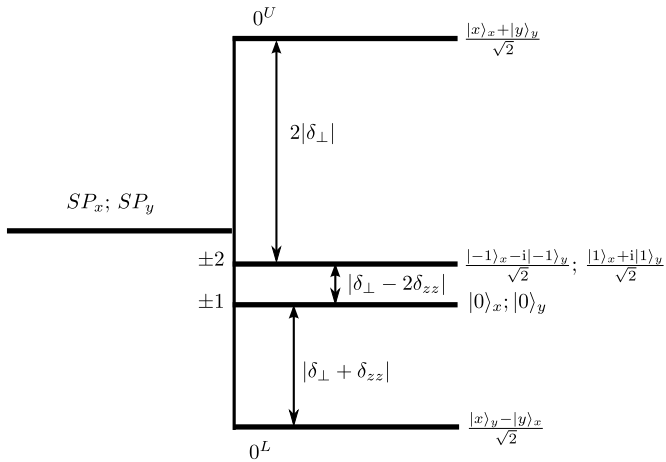}
\caption{Schematic illustration of the fine structure of triplet states in an isotropic quantum disk. $SP_x$ and $SP_y$ denote degenerate orbital states $\Psi_x(\bm \rho_1,\bm \rho_2)$ and $\Psi_x(\bm \rho_1,\bm \rho_2)$ [Eq.~\eqref{psifun}], respectively. $|0\rangle$, $|+1\rangle$, $|-1\rangle$, $|x\rangle$ and $|y\rangle$ denote spin states. Subscripts $x$ and $y$ denote orbital states. $0^L$, $0^U$, $\pm 1$ and $\pm 2$ label total momentum $z$ component.}
\label{fig:iso} 
\end{figure}

In \emph{isotropic} quantum disks the orbitals $SP_x$ and $SP_y$ are degenerate. The rotation by $\pi/2$ around $z$-axis transforms $SP_x$ state into $SP_y$ state. Therefore, they cannot be considered separately and effective spin-dependent Hamiltonian contains both diagonal in orbital states part, $\hat \Delta_{ii}$, and non-diagonal parts, $\hat\Delta_{ij}$ ($i\ne j$). The diagonal part is given by Eq.~\eqref{H:sym:ii} with $\mathcal A_x = \mathcal B_y$, $\mathcal A_y = \mathcal B_x$. Non-diagonal in orbital indices contributions read
\begin{equation}
 \label{H:sym:yx}
\hat{\Delta}_{xy} = \hat{\Delta}_{yx} = (\mathcal A_x - \mathcal B_x) \{\hat S_x,\hat S_y\}_{\rm sym},
\end{equation}
where $\{\hat A,\hat B\}_{\rm sym} = (\hat A \hat B+\hat B\hat A)/2$. 

Schematic level structure is shown in Fig.~\ref{fig:iso}. The states can be classified in accordance to their total momentum (spin and angular) $z$-component $F_z$ and form two non-degenerate sublevels: $0^L$ and $0^U$ with $F_z=0$ and two two-fold degenerate sublevels with $F_z = \pm 1$ and $\pm 2$. The fine structure of two-electron complex in isotropic disks is analogous to that of excited exciton states in isotropic quantum disks, cf.~\cite{glazov2007a}.

\subsection{Microscopic theory}

Each triplet described by an antisymmetric orbital wavefunction
\begin{equation}\label{psifun}
 \Psi_{i}(\bm \rho_1,\bm \rho_2) =\frac{1}{\sqrt{2}} \left[S(\bm \rho_1) P_i(\bm \rho_2) - S(\bm \rho_2) P_i(\bm \rho_1)\right]
\end{equation}
is lower in energy as compared with the corresponding singlet, because the Coulomb interaction for this state is weaker due to the smaller overlap of electron wavefunctions in the triplet as compared with the singlet. The singlet-triplet splitting is of the order of $e^2/\varkappa a$ and is much larger than the fine structure splittings of triplet state (to be calculated below). Therefore, one can neglect possible admixture of a singlet state due to the spin-orbit interaction.

The effective Hamiltonian describing a fine structure of a pair of carriers can be obtained by taking the matrix elements of Eq.~\eqref{Coulomb1} in the basis of triplet $SP_i$. $SP_j$ states. It turns out that the direct contribution vanishes and the exchange one can be written as
\begin{equation}\label{Delta}
\hat{ \Delta}_{ij}= - \frac{2\pi e^2\xi^2}{\Xi\varkappa} \sum_{\bm k, \bm k', \bm p, \bm p'}  \frac{\delta_{\bm k+ \bm k', \bm p + \bm p'}}{|\bm k - \bm p|} \tilde S^*(\bm p') \tilde P^*_j(\bm p) \tilde S(\bm k) \tilde P_i(\bm k') 
\end{equation}
\[
\{F_{11}^{11}(q)[\hat{\bm \sigma}^{(1)} \times (\bm p+\bm k)]_z [\hat{\bm \sigma}^{(2)} \times (\bm p'+\bm k')]_z
\]
\[
 -F_{22}^{00}(q) [\bm p \times \bm k]_z [\bm p' \times \bm k']_z \hat{ \sigma}^{(1)}_z \hat{ \sigma}^{(2)}_z \}, \quad i,j=x\mbox{ or }y.
\]
Here the Fourier transform of an envelope function, e.g. of $S(\bm \rho)$, is defined as
\[
 \tilde S(\bm k) = \frac{1}{\sqrt{\Xi}}\int e^{-\mathrm i \bm k \bm \rho}\: S(\bm \rho) \mathrm d \bm \rho.
\]
It is worth noting that the Hamiltonian Eq.~\eqref{Delta} does not describe small spin-orbit induced corrections to the singlet-triplet splitting of electron states in quantum dots. In contrast with the exchange interaction of an electron and a hole in an exciton which is screened by a high-frequency dielectric constant, see Refs.~\cite{ivchenko05a,goupalov98,glazov2007a} and references therein, both direct and exchange interaction of two electrons are screened by the low-frequency dielectric constant because all involved energies are small as compared with the band gap $E_g$~\cite{ivchenko05a,badescu:161304,ee}.

\textit{Anisotropic quantum disk.} We assume that the quantum disk anisotropy is small enough that the deformation of the wavefunctions is negligible and $SP_x$ orbital transforms to $SP_y$ orbital by a rotation by $\pi/2$. The energy splitting between these orbitals $\varepsilon$, however, is supposed to exceed by far the fine structure splittings of each triplet. One can consider $SP_x$ and $SP_y$ orbitals as independent but neglect the differences in their shapes a wide range of $\varepsilon$ values, because the interlevel splitting in the quantum disk (determined by the confinement) is much larger than the fine structure splittings of each triplet state (determined by the spin-orbit interaction). One can check that in this case
\begin{equation}
 \label{Delta:qd:1}
\mathcal A_x = \mathcal B_y = - \frac{2}{3}(\delta_\perp + \delta_{zz}), \quad \mathcal B_x = \mathcal A_y=  \frac{2}{3}(2\delta_\perp - \delta_{zz}),
\end{equation}
where we introduced new parameters $\delta_\perp$ and $\delta_{zz}$ describing the splittings between linearly polarized states and between the state with $m_z=0$ and one of the linearly-polarized states, respectively, in agreement with Fig.~\ref{fig:aniso}.

The calculation shows that for the parabolic quantum disk where the electron pair states are described by Gaussian envelopes such as
\[
 S(\bm \rho) =  \frac{1}{\sqrt{2\pi a^2}} e^{-\rho^2/4a^2} , \; P_x(\bm \rho) = \frac{x}{a} S(\bm \rho), \; P_y(\bm \rho) = \frac{y}{a} S(\bm \rho),
\]
where $a$ is an effective disk radius the constants $\delta_{\perp}$ and $\delta_{zz}$ have the following form
\begin{equation}
 \label{deltas}
\delta_{\perp} =  -\xi^2\frac{3 e^2}{8 \varkappa da^4},\quad
\delta_{zz} = - \xi^2 \frac{\sqrt{\pi}e^2}{32 \varkappa a^5}.
\end{equation}
Here $d$ is the height of the quantum disk, the infinite barriers are assumed to describe quantization along $z$-axis. Interestingly, $\delta_{\perp}$ is parametrically larger than $\delta_{zz}$, i.e. their ratio $\delta_{\perp}/\delta_{zz} \sim a/d> 1$ in quantum disks. Therefore, as it is shown in Fig.~\ref{fig:aniso} the splitting between $m_z=0$ and one of linearly-polarized states is substantially smaller than the energy distance to another one.

\begin{figure}
\includegraphics[width=0.9\linewidth]{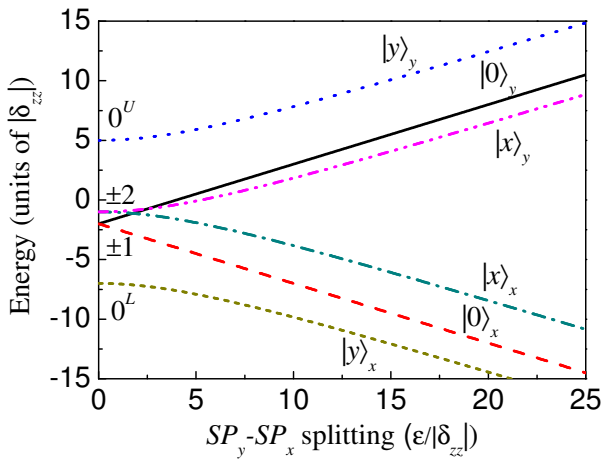}
\caption{(Color online). Fine structure of $SP$ triplet states as a function of quantum disk anisotropy. Levels are denoted in the same way as in Figs.~\ref{fig:aniso},~\ref{fig:iso}. The case of $\delta_{\perp}/\delta_{zz} =3$ is shown.}
\label{fig:isoaniso} 
\end{figure}

\textit{Isotropic quantum disk.} In quantum disks with the axially-symmetric or square-shaped lateral potential the two orbital states $SP_x$ and $SP_y$ can not be considered as independent ones and the non-diagonal in orbital indices matrix elements ($i\ne j$) of the $\hat{\Delta}_{ij}$ spin Hamiltonian [see Eq.~\eqref{H:sym:yx}] should be taken into account. In agreement with the symmetry of the system there are two doubly-degenerate states with the total momentum $z$-component $F_z = \pm 2$ and $F_z =\pm 1$ and two non-degenerate states with the total momentum component being $0$, see Fig.~\ref{fig:iso}.

The transition between the cases of isotropic and anisotropic quantum disks takes place where the splitting between $SP_x$ and $SP_y$ orbitals $\varepsilon$ becomes comparable with $\delta_{\perp}$. The level arrangement in the case of arbitrary quantum disk anisotropy (provided $\varepsilon$ is much smaller than the distance between $SS$ and $SP$ levels so that the deformation of the wavefunctions can be neglected) is shown in Fig.~\ref{fig:isoaniso}.

\section{Discussion and conclusion}

The ratios of the fine structure splitting constants $\delta_\perp$, $\delta_{zz}$ and the Coulomb exchange interaction energy of an electron pair in the $SP$ orbital state are the important dimensionless parameters which determine the relative magnitudes of the triplet fine structure splittings. The Coulomb exchange interaction energy equals to the half of the singlet-triplet splitting and reads
\[
U_e = \frac{e^2}{\varkappa} \int \frac{\mathrm d \bm \rho_1 \mathrm d\bm \rho_2}{|\bm \rho_1 - \bm \rho_2|} S(\bm \rho_1) P_x(\bm \rho_1)P_x(\bm \rho_2)S(\bm \rho_2) = \frac{\sqrt{\pi} e^2}{8\varkappa a}.
 \]
Therefore
\begin{equation}
 \left|\frac{\delta_{\perp}}{U_e}\right| = \frac{3\xi^2}{\sqrt{\pi}\, da^3}, \quad \left|\frac{\delta_{zz}}{U_e}\right| = \frac{\xi^2}{4a^4}.
\end{equation}
The parameter $|\xi|\approx 5$~\AA$^2$ (GaAs), $30$~\AA$^2$ (GaSb), $110$~\AA$^2$ (InAs) and $500$~\AA$^2$ (InSb)~\cite{PhysRevB.55.16293}. The splittings are therefore quite small in GaAs based quantum dots but can be strongly enhanced in narrow-band semiconductor materials. These splittings of the triplet state can be observed in polarization-resolved spectroscopy of the doubly-charged quantum dot where, after the recombination of an additional electron-hole pair injected into the dot, a pair of electrons occupying $S$ and $P$ orbitals remains.

In agreement with the symmetry considerations the fine structure of triplet states arises even in the centrosymmetric systems. An effective point symmetry of isotropic and anisotropic disks considered here is $D_{4h}$ and $D_{2h}$, respectively, as compared with $D_{2d}$ and $C_{2v}$ point groups describing real III-V semiconductor quantum dots. However, an absence of an inversion center in the symmetry group of either bulk material or of heteropotential does not change qualitatively the fine structure of triplet states. 

The case of non-centrosymmetric system has been analyzed in a number of works, see Refs.~\cite{kkavokin01,kkavokin04,gangadharaiah:156402}. In those papers the conduction band spin splitting due to bulk or structure inversion asymmetry was taken into account but the spin-orbit corrections to the electron-electron interaction were disregarded. It was demonstrated that fine structure of the triplet state is absent in the first order in conduction band spin splittings. In that case, effectively, the exchange interaction of electrons remains the same, but the spins of carriers are rotated due to the effective magnetic field arising from the spin splitting. The complete removal of spin degeneracy due to the conduction band spin splitting was demonstrated in the \emph{forth} order in the spin-orbit coupling~\cite{gangadharaiah:156402} while the present mechanism gives rise to the fine structure in the \emph{second} order in spin-orbit coupling. 

In conclusion, we have derived the spin-orbit terms in the effective matrix element of electron-electron interaction in quantum wells. We have addressed theoretically the fine structure of two electron complexes localized in small lateral quantum dots. We have shown that the spin degeneracy of the two electron triplet states is completely lifted in anisotropic quantum dots. In quantum disks with isotropic lateral potential the two electron states are split in accordance with the total angular momentum projection. The fine structure splittings are calculated for the quantum disks with the parabolic lateral potential.

We note that the exact values of the triplet state splittings depend strongly on quantum disk parameters and on interface properties. The extension of the present theory to the realistic quantum dot systems (including double dots, vertically coupled quantum dots and semiconductor nanocrystals of an arbitrary shape) may be a possible development of this study. Another steps to continue this investigation are: (i) take into account an absence of an inversion center in bulk material and analyze an interference of bulk and structure asymmetry in the triplet state fine structure, (ii) study an interplay of conduction band spin splitting and spin-orbit terms in electron-electron interaction in the formation of fine structure of a two-electron state in a quantum dot.

\acknowledgements

We are grateful to E.L. Ivchenko and K.V. Kavokin for useful discussions. The financial support of RFBR, Programmes of RAS and the ``Dynasty'' Foundation -- ICFPM is acknowledged.

\end{document}